\documentclass[aps,prc,preprint,superscriptaddress,showpacs,floatfix]{revtex4}

\usepackage{graphicx}
\usepackage{longtable}
\usepackage{bm}
\usepackage{color}

\begin{document}

\title{ No evidence of an 11.16 MeV 2$^{+}$ state in $^{12}$C}

% ---
\author{F.~D.~Smit}
\email[Email address: ]{smit@tlabs.ac.za}
\affiliation{iThemba Laboratory for Accelerator Based Sciences, Somerset West 7129, South Africa}
% ---
\author{F.~Nemulodi}
\affiliation{iThemba Laboratory for Accelerator Based Sciences, Somerset West 7129, South Africa}
\affiliation{Department of Physics, University of Stellenbosch, Matieland 7602, South Africa}
% ---
\author{Z.~Buthelezi}
\affiliation{iThemba Laboratory for Accelerator Based Sciences, Somerset West 7129, South Africa}
% ---
\author{ J.~Carter}
\affiliation{School of Physics, University of the Witwatersrand, Johannesburg 2050, South Africa}
% ---
\author{ R.F.~Fearick}
\affiliation{Department of Physics, University of Cape Town, Rondebosch 7700, South Africa}
% ---
\author{ S.~V.~F\"ortsch}
\affiliation{iThemba Laboratory for Accelerator Based Sciences, Somerset West 7129, South Africa}
% ---
\author{M.~Freer}
\affiliation{School of Physics and Astronomy, University of Birmingham, Birmingham B15 2TT, United Kingdom}
% ---
\author{H.~Fujita}
\affiliation{Research Center for Nuclear Physics, Osaka University, Ibaraki, Osaka 567 0047, Japan}
% ---
\author{M.~Jingo}
\affiliation{School of Physics, University of the Witwatersrand, Johannesburg 2050, South Africa}
% ---
\author{C.~O.~Kureba}
\affiliation{School of Physics, University of the Witwatersrand, Johannesburg 2050, South Africa}
% ---
\author{J.~Mabiala}
\affiliation{Department of Physics, University of Stellenbosch, Matieland 7602, South Africa}
% ---
\author{J.~Mira}
\affiliation{iThemba Laboratory for Accelerator Based Sciences, Somerset West 7129, South Africa}
% ---
\author{R.~Neveling}
\affiliation{iThemba Laboratory for Accelerator Based Sciences, Somerset West 7129, South Africa}
% ---
\author{P.~Papka}
\affiliation{Department of Physics, University of Stellenbosch, Matieland 7602, South Africa}
% ---
\author{G.~F.~Steyn}
\affiliation{iThemba Laboratory for Accelerator Based Sciences, Somerset West 7129, South Africa}
% ---
\author{J.~A.~Swartz}
\affiliation{iThemba Laboratory for Accelerator Based Sciences, Somerset West 7129, South Africa}
\affiliation{Department of Physics, University of Stellenbosch, Matieland 7602, South Africa}
% ---
\author{I.~T.~Usman}
\affiliation{iThemba Laboratory for Accelerator Based Sciences, Somerset West 7129, South Africa}
% ---
\author{J.~J.~van~Zyl}
\affiliation{Department of Physics, University of Stellenbosch, Matieland 7602, South Africa}
% ---

\date{\today}

\begin{abstract}
An experiment using the $^{11}$B($^{3}$He,d)$^{12}$C reaction was performed at iThemba LABS at
an incident energy of 44 MeV and analyzed with a high energy-resolution magnetic spectrometer, to
re-investigate states in $^{12}$C  published in 1971.  The original investigation reported the existence
of an 11.16 MeV state in $^{12}$C that displays a 2$^{+}$ nature. In the present experiment data were
acquired at laboratory angles of 25$^{\circ}$, 30$^{\circ}$ and 35$^{\circ}$, to be as close to the c.m.
angles of the original measurements where the clearest signature of such a state was observed. These
new low background measurements revealed no evidence of the previously reported state at 11.16 MeV
in $^{12}$C.
\end{abstract}



\maketitle

%---
% INTRODUCTION
%---

The $^{12}$C nucleus, through the triple-alpha configured 0$^{+}$ state at 7.65 MeV (known as the
Hoyle state), provides the gateway for the synthesis of all elements from itself to the heaviest elements
occurring naturally \cite{Fyn05}. Locating the 2$^{+}$ excited state associated with this 0$^{+}$ state in
$^{12}$C will provide information of a structural nature about the triple-alpha configuration of this
0$^{+}$ state. Of even greater importance, cluster calculations \cite{Des87} predict the existence
of this associated 2$^{+}$ state at 9.11 MeV. On the strength of these calculations it has been included
in the European based astrophysical nucleosynthesis calculations, NACRE \cite{Ang99}. Here it
contributes to the formation of $^{12}$C through helium burning at high temperatures, as experienced
in Type II supernovae.

Longstanding published data \cite{Rey71} do exist where a state with a suggested 2$^{+}$ nature was seen,
however, at 11.16 MeV, but that is presently still unconfirmed. These data are from an experiment performed
at an incident energy of 44 MeV, using the $^{11}$B($^{3}$He,d)$^{12}$C reaction on an enriched $^{11}$B
target, with the reaction particles analysed using a double-focusing magnetic spectrometer equipped with
photographic plates at the focal plane.  For many years this has been the only existing candidate in the data
for the first excited state of the Hoyle state.

Interest in the 2$^{+}$ state gained new momentum since the ($\alpha$,$\alpha$$^\prime$) data of Itoh {\it et al.}
\cite{Ito04} was first published in 2004, in which it was suggested that a weak 2$^{+}$ state exists in $^{12}$C
at E$_{\it x}$ $\sim$ 10 MeV, where it is partially obscured by the other strong states in the excitation energy
region of 9 - 11 MeV. High energy-resolution (p,p$^\prime$) data at 66 and 200 MeV \cite{Fre09} acquired at
iThemba LABS confirmed that a broad state with a 2$^{+}$ character exists at 9.6(1) MeV, buried beneath the very
strong 3$^{-}$ state at 9.64 MeV. Recently, Itoh {\it et al.} \cite{Ito11} presented a refined analysis of their
($\alpha$,$\alpha$$^\prime$) study at 386 MeV, using both peak-fitting and multipole decomposition techniques.
They concluded that a 2$^{+}$ state with a width of 1.01 $\pm$ 0.15 MeV exists at E$_{\it x}$ = 9.84 $\pm$ 0.06 MeV.
Further support for the existence of a broad state with a 2$^{+}$ character in the region of 9.6 MeV also came
from a study of low energy (25 MeV) inelastic proton scattering by Zimmerman et al. \cite{Zim11}, as well as a
$^{12}$C($\gamma$,3$\alpha$) experiment by Gai {\it et al.} \cite{Gai11}.  Gai {\it et al.} explicitly state
that no evidence of a state at 11.16 MeV was found. Recently, work was published \cite{Kir10} where the
$^{11}$B($^{3}$He,d)$^{12}$C reaction at an incident energy of 8.5 MeV was used to investigate $^{12}$C
breakup into three $\alpha$-particles. Although not explicitly mentioned, no strength for a state at
11.16 MeV can be seen in Fig. 6 of that paper \cite{Kir10}.

In view of the great interest in the latest evidence on the location of the 2$^{+}$ state, as well as even a
possible 4$^{+}$ state at 13.3 MeV \cite{Fre11}, it became necessary to revisit the measurement of
Ref.\cite{Rey71} for a more rigorous investigation with the improved equipment and analysis techniques
available today.

%
%EXPERIMENT AND ANALYSIS
%
The repeat of the $^{11}$B($^{3}$He,d)$^{12}$C measurements (Q = +10.463 MeV) reported here was performed
at iThemba LABS, South Africa at an incident beam energy of 44 MeV, as in the case of the measurements by
Reynolds {\it et al.} \cite{Rey71}. Data were acquired at three laboratory angles over the course of two and a
half days. The present experiment was performed with the K600 magnetic spectrometer \cite{Nev11}. The
emitted deuterons were detected with a focal-plane detector system that consisted of two multi-wire drift
chambers (MWDC) and a plastic scintillator. Modern electronics and computers allow for a far more rigorous
analysis of the data than was possible previously with photographic plates.

The target used for this experiment was a self supporting $^{11}$B foil with an areal density of
395 $\mu$gcm$^{-2}$ made from 98$\%$ enriched material. In addition, a self supporting 600 $\mu$gcm$^{-2}$
natural boron target was also used to investigate the influence of possible contamination from the remnant
$^{10}$B in the enriched target. In the paper on the original measurements \cite{Rey71} it states that their
target was enriched, but gave no further details.

An Elastic Recoil Detection Analysis (ERDA) was performed on the enriched $^{11}$B target to determine the
extent of the contaminants on the target. The target contained approximately 212 $\mu$gcm$^{-2}$ of $^{11}$B
as well as strong $^{16}$O contamination of approximately 170 $\mu$gcm$^{-2}$ (Q = -4.893 MeV)
and to a lesser extent $^{12}$C with a thickness of approximately 6 $\mu$gcm$^{-2}$ (Q = -3.550 MeV). Although
the oxygen contamination of the target was strong, the Q-value difference between the reactions on
$^{11}$B and $^{16}$O moved contamination peaks out of the energy region of interest. Weak $^{14}$N
contamination, approximately 7 $\mu$gcm$^{-2}$ (Q = +1.803 MeV), was also measured. Nitrogen is believed to
be found in boron nitride compounds formed with the residual gas during the vacuum evaporation process.

In the present experiment, data were taken at three laboratory angles, namely 25$^{\circ}$, 30$^{\circ}$
and 35$^{\circ}$. It was at these larger angles in the original experiment where the signature for a
2$^{+}$ state was recorded \cite{Rey71}. A further advantage of this choice of angles was that a
well-shielded external beam stop some 10 m away from the scattering chamber could be used. At smaller
angles, an internal beam stop inside the scattering chamber would have been required that could have led
to unwanted background.

%%%%%%% Figure1 %%%%%%%%%%%%%%%
\begin{figure}
\includegraphics[scale=0.6,angle=0]{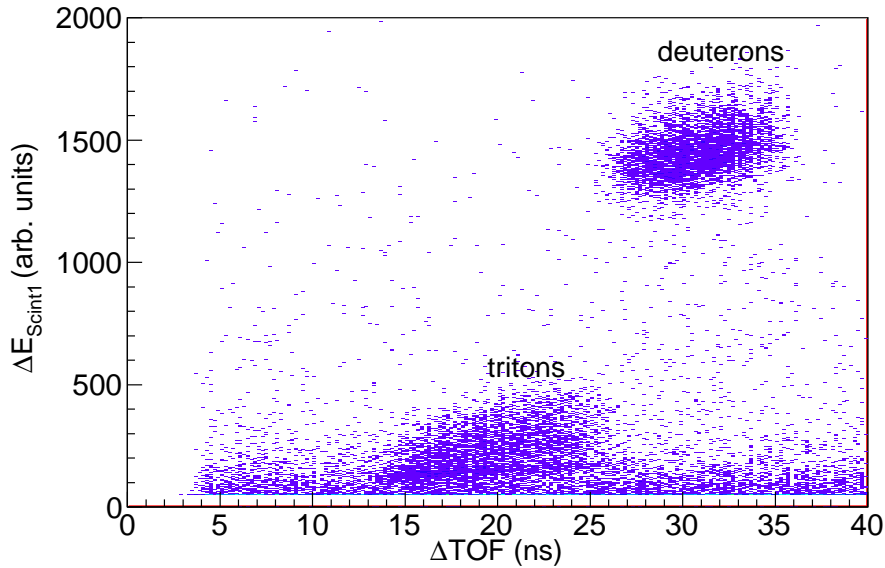}
\caption{\label{fig1}A particle identification spectrum of the pulse height in the scintillator {\it versus} the
time-of-flight (ToF) through the spectrometer. Deuterons and tritons are clearly separated. }
\end{figure}
%%%%%%%%%%%%%%%%%%%%%%%%%%%%%%

At the dipole magnetic field settings required for the experiment only deuterons and tritons have the required
magnetic rigidity to reach the focal plane detectors. As can be seen in Fig. \ref{fig1}, the deuteron and triton
signatures were well separated. The vertical axis of the spectrum represents energy deposited in the scintillator
while the horizontal axis is the time-of-flight (ToF) of the particles through the spectrometer measured
between the Separated Sector Cyclotron radio-frequency (RF) signal and the scintillator trigger. In the off-line
analysis a software gate was set to include only the deuterons. Empty frame runs were also made to characterize
possible beam background contamination, but only a negligible number of events were observed in the deuteron gate.

%%%%%%% Figure2 %%%%%%%%%%%%%%%
\begin{figure}
\includegraphics[scale=0.6,angle=0]{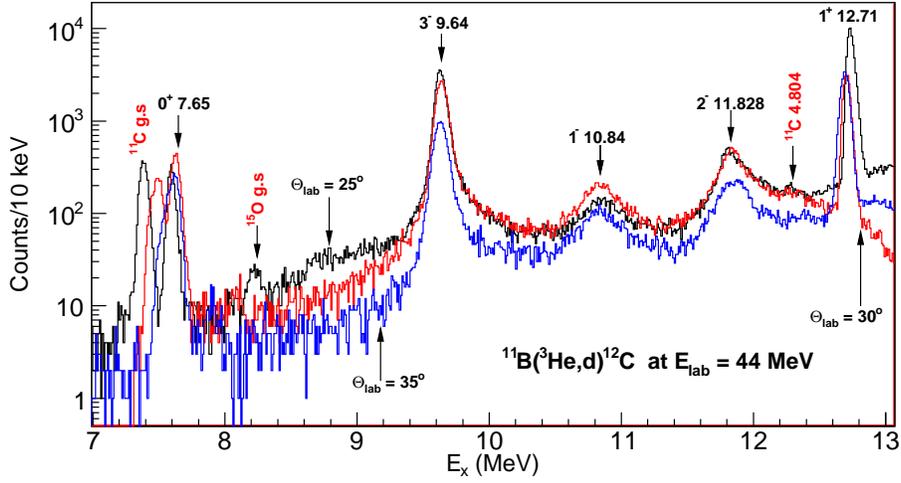}
\caption{\label{fig2}Excitation energy spectra for the $^{11}$B($^{3}$He,d)$^{12}$C reaction at laboratory
emission angles of 25$^{\circ}$ (black),  30$^{\circ}$ (red) and 35$^{\circ}$ (blue). (color online) }
\end{figure}
%%%%%%%%%%%%%%%%%%%%%%%%%%%%%%

Figure \ref{fig2} shows the $^{12}$C excitation energy spectra measured at the three emission angles. Note that there
is a deep valley where the 11.16 MeV state is expected. In the original paper \cite{Rey71}, a broad state with
strength almost equal to that of the 10.84 MeV state can be seen filling in the valley between the 10.84 MeV state
and the 11.83 MeV state. No such strength is indicated at 11.16 MeV in the present data set.

%%%%%%% Figure3 %%%%%%%%%%%%%%%
\begin{figure}
\includegraphics[scale=0.6,angle=0]{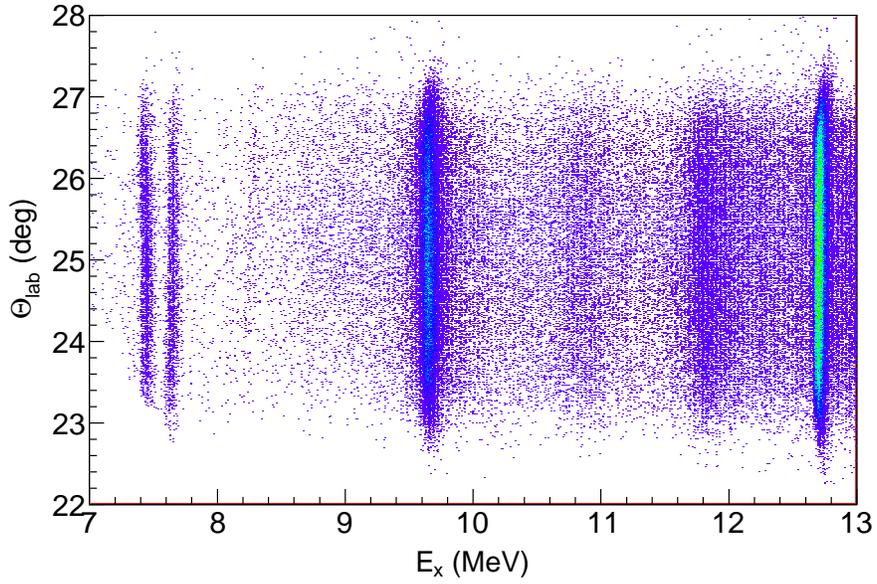}
\caption{\label{fig3}A two dimensional plot of the deuteron emission angle {\it versus} $^{12}$C excitation energy
for $\Theta$$_{lab}$ = 25$^{\circ}$}
\end{figure}
%%%%%%%%%%%%%%%%%%%%%%%%%%%%%%

The two dimensional plot in Fig. \ref{fig3} of deuteron emission angle versus $^{12}$C excitation energy clearly indicates
 the capabilities of modern equipment and analysis techniques. Here, the spectrometer settings and data analysis
have been optimized so that the $^{11}$B($^{3}$He,d)$^{12}$C reaction appears as vertical loci while the
($^{3}$He,d) reaction on heavier (e.g. see $^{15}$O) target nuclei slope forward and  conversely for lighter
(e.g. see $^{11}$C) target nuclei, slope backward. The projection of this spectrum onto the horizontal axis yields
the one dimensional excitation energy spectrum of the data at 25$^{\circ}$, as seen in Fig. \ref{fig2}. Such an overview
assists in determining whether a broad peak in the one dimensional spectrum belongs to the nucleus of interest,
or if it is from some other contaminant nucleus with a different kinematic correction resulting in a broad peak.
A good example of this is the $^{15}$O ground state which appears in the spectra at around 8.3 MeV excitation
in Fig.\ref{fig2} , and can be seen in Fig. \ref{fig3} to be a locus with a slope. Thus, Fig. \ref{fig3} also allows one to form a visual
impression of the extent of the background contamination.

%%%%%%% Figure4 %%%%%%%%%%%%%%%
\begin{figure}
\includegraphics[scale=0.5,angle=0]{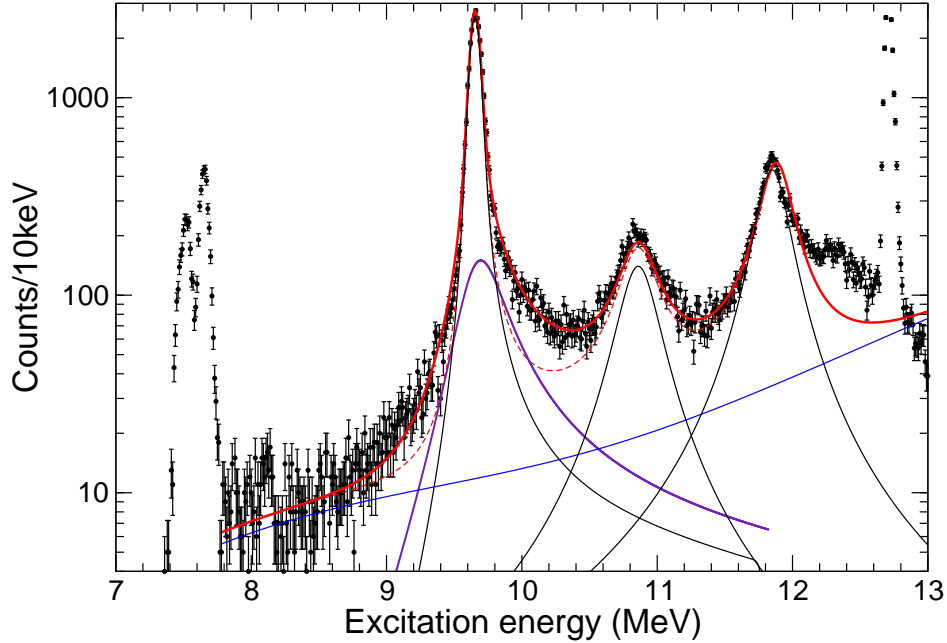}
\caption{\label{fig4}Excitation energy spectrum at $\Theta_{lab}$ = 30$^{\circ}$ with the peaks at 10.84 MeV and
11.83 MeV fitted with a Lorentzian line-shape, with $\Gamma$ = 315 keV and $\Gamma$ = 260 keV, respectively.
At 9.64 MeV the 3$^{-}$ peak was generated from an R-matrix calculation convoluted with the 65 keV experimental
energy resolution. Added to that is a 2$^+$ resonance peak at 9.7 MeV also generated from an R-matrix calculation.
Here, the blue curve is a background with the solid red curve an overall fit with, and the dashed red curve without a
2$^+$ state at 9.7 MeV. (colour online)}
\end{figure}
%%%%%%%%%%%%%%%%%%%%%%%%%%%%%%

By way of example, shown in Fig. \ref{fig4}, are fits to the peaks of the 30$^{\circ}$ data. The sum of all fits is indicated
by the solid red curve. Good overall agreement is achieved to data in the region around 11.16 MeV excitation,
indicating that no other strength is required in this energy region. It also indicates the desirability of adding a
peak at an excitation energy of 9.7 MeV to achieve a good fit in this region.

An attempt was made to determine the origin of the 11.16 MeV peak seen in the original data \cite{Rey71}.
As far as could be ascertained from the data acquired on the $^{11}$B and $^{nat}$B targets, and knowing
the contaminants on these targets, as well as investigating tritons as the possible emitted reaction particle,
no clear cause could be found for the signature seen in the original experiment.

%
%CONCLUSIONS
%
The $^{11}$B($^{3}$He,d)$^{12}$C measurement at an incident energy of 44 MeV of Reynolds {\it et al.}
was repeated with a high energy-resolution magnetic spectrometer and no evidence was found for
the previously reported 2$^{+}$ state at 11.16 MeV in $^{12}$C. In a 4 - 5 MeV excitation energy
region above the Hoyle state at 7.65 MeV, the only remaining 2$^{+}$ state is the state reported to be
in the region of 9.6 - 9.8 MeV. There is now a need for theoretical cluster calculations to investigate
this state as a possible member of a band associated with the Hoyle state. Theoretical studies of the
astrophysical implications of the energy and strength of this 2$^{+}$, as well as the newly identified
4$^{+}$ state at 13.3 MeV in $^{12}$C should also prove to be very interesting.

%
%ACKNOWLEDGMENTS
%
The authors would like to thank Dr Mandla Msimanga of iThemba LABS (Gauteng) for the ERDA
measurements on the targets, as well as Dr Lowrie Conradie and his accelerator team for providing
the excellent beam during the experiment. This work was supported by the South African National
Research Foundation.

%%%%%%%%%%%%%%%%%%%%%%%%%%%%%%%%%%%%%%%%%%%%%%%%%%%%%%%%%%%%%%%%%%
%%%%% references %%%%%

\end{document}